\documentclass[prl,twocolumn, amsmath,amssymb,
reprint,%
author-year,%
author-numerical,%
dvipdfmx,
superscriptaddress
]{revtex4-1}

\usepackage{graphicx}
\usepackage{bm}
\usepackage{color}

\begin{document}

\title{A Framework for Identifying Non-van der Waals 2D Materials}

\author{Shota Ono}
\email{shotaono@muroran-it.ac.jp}
\affiliation{Department of Sciences and Informatics, Muroran Institute of Technology, Muroran 050-8585, Japan}

\begin{abstract}
Two-dimensional (2D) materials are categorized into van der Waals (vdW) and non-vdW types. 
However, no relevant descriptors have been proposed for identifying the latter. 
Here, we identify the non-vdW 2D materials by calculating the thickness-dependence of total energy of thin films truncated from surfaces. The non-vdW 2D materials exhibit a deviation from the law of exfoliation energy inverse to the number of layers in the monolayer limit. This framework is applied to explore single- and multi-component systems, which predicts the synthesizability of several non-vdW 2D materials including silicene and goldene that are overlooked in the dimensional analysis of the parent crystals and also predicts that a Janus structure exists in nature but is hidden in 3D crystals.    
\end{abstract}

\maketitle

{\it Introduction---}Since the successful exfoliation of graphene from graphite \cite{graphene}, other two-dimensional (2D) materials that exhibit interesting properties have been explored. Several geometric approaches have been used to identify a few thousands of layered materials among several material databases. As the pioneering work, Leb\`{e}gue {\it et al}. identified about 100 van der Waals (vdW) materials that have a large interlayer distance along the $c$ axis \cite{lebegue}. Ashton {\it et al}. developed a topology-scaling algorithm determining the dimensionality of the parent materials and identified 800 candidate materials in the Materials Project (MP) database \cite{ashton}. Mounet {\it et al}. also identified about 1000 vdW materials using the Inorganic Crystal Structure Database (ICSD) and the Crystallographic Open Database (COD) \cite{mounet}. Given the 2D structure prototypes, the element substitutions are employed to expand the family of 2D materials systematically, which led to the development of the Computational 2D materials Database (C2DB) \cite{C2DB2021}. More recently, a machine learning algorithm has been used to generate 2D structures that are different from the known prototypes \cite{lyngby}. Furthermore, synthetic 2D structures (e.g., MoSi$_2$N$_4$ \cite{hong}) having no 3D counterparts have been extensively investigated. 

Although their parent materials are classified into non-vdW (non-layered) materials, several 2D materials have been synthesized experimentally, such as Xenes \cite{review,kashiwayareview} including silicene \cite{vogt}, plumbene \cite{yuhara}, and goldene \cite{kashiwaya}, and binary and ternary 2D compounds \cite{ji2019,puthirath2021,balan2022}. Therefore, more than tens of thousands of non-vdW 2D candidate materials have been overlooked in constructing 2D materials database. Basically, non-vdW 2D materials have been explored by truncating a monolayer from a specific surface of the 3D counterparts, and their exfoliation and cohesive energies and dynamical and thermodynamic stabilities have been investigated \cite{2Dtraditional,nevalaita2018,ono2020,pawar2022,friedrich2022,ono2024,pereira2025}. However, predicting the synthesizability of non-vdW 2D materials \cite{ono2021}, as well as that of 3D compounds \cite{sun} and 0D clusters \cite{de}, is still under debate. In this Letter, we provide an approach for identifying non-vdW 2D materials based on the energy variation of thin films in the monolayer limit.

So far, vdW 2D materials have provided a fertile platform to explore interesting properties such as magnetism \cite{pakdel}, superconductivity \cite{wines}, and heavy-fermion state \cite{jang}. The 2D semiconductors have also been explored for next-generation electronics and optoelectronics \cite{wang,zhang}. The present work paves the way for the physics and their interesting applications based on the non-vdW 2D materials. 

{\it Basic idea---}Let us consider the carbon, silicon, and germanium crystals in the diamond structure and truncate thin films from the (111) surface. If the film thickness is decreased, the electronic character in the occupied states changes from $sp^3$ to $sp^2$ and the thin films are transformed into multi-layered graphene, silicene, and germanene at a specific thickness. This is a hypothetical creation of 2D materials from non-vdW materials. Therefore, it is important to study the energy variation of thin films for exploring non-vdW 2D materials, which may reflect such a 3D-2D transformation.

For a quantitative discussion, we define the exfoliation energy as $\Delta E(N) = E(N)/N-E_{\rm bulk}$, where $E(N)$ and $E_{\rm bulk}$ are the total energy per unit cell of $N$-layer thin film and bulk, respectively. 

{\it Theorem: $\Delta E \propto N^{-1}$ for large $N$.} 

{\it Proof:} We assume that the surface relaxation occurs only at the top and bottom layers. The $N$ layer thin film consists of $(N-2)$ bulk layers and two surface layers, where the chemical bonds on the surface are different from those in the bulk. We can express the total energy of this system as $E(N) \simeq (N-2)E_{\rm bulk}+2E_{\rm surf}$, where $E_{\rm surf}$ is the energy of the surface layer for large $N$. Thus, one obtains $\Delta E(N) = E_0/N$, where the factor $E_0=2(E_{\rm surf} - E_{\rm bulk})$ is independent of $N$. 

On the contrary, $\Delta E(N=1)=E_{\rm mono}- E_{\rm bulk}$ with $E_{\rm mono}$ being the total energy of monolayer. If $E_{\rm mono}$ is equal to $-E_{\rm bulk}+2E_{\rm surf}$, the $N^{-1}$ law exactly holds. If a parent non-vdW material can take the 2D form as a metastable structure, $E_{\rm mono}$ is significantly lowered. In this case, one will observe a breakdown of the $N^{-1}$ law. In this Letter, we propose that ``a downward deviation from the $N^{-1}$ scaling of $\Delta E$ at $N=1$'' is a good indicator to identify non-vdW 2D materials. Our approach is simply based on the energy calculations of thin films, avoiding the ambiguous argument about the chemical bonds. 

When the $N^{-1}$ law holds, one obtains $\ln \Delta E(N) = - \ln N + C$ with a constant $C$. To identify the deviation from the $N^{-1}$ law, we calculate the following quantity: 
\begin{eqnarray}
p(N) = \frac{d \ln \Delta E}{d\ln N} \simeq 
\frac{\ln \Delta E(N+1) - \ln \Delta E(N)} { \ln (N+1) - \ln N }.
\label{eq:pN}
\end{eqnarray}
If $p(N)=-1$, the $N$-layer thin film exactly follows the $N^{-1}$ law. The inequality of $p(1)>-1$ indicates a downward deviation from the $N^{-1}$ law. 

{\it Computational details---}We used Quantum Espresso \cite{qe} to perform density-functional theory (DFT) calculations and used the Perdew-Burke-Ernzerhof (PBE) functional \cite{pbe} in the pslibrary1.0.0 \cite{dalcorso}. The $4f$ states of lanthanides were treated as the core states. The energy cutoff for wavefunction $E_{\rm cut}$ was set to be the highest $E_j$ plus 10 Ry, where $E_j$ is the suggested value for atom $j$ in each compound, and the energy cutoff for charge density was set to be $10E_{\rm cut}$. The $k$ point distance of $\Delta k \le 0.15$ \AA \ was assumed. The smearing parameters of 0.015 and 0.01 Ry were used for single- and multi-component systems, respectively \cite{smearingMV}. The vacuum layer was set to be 20 \AA \ to avoid the spurious interactions between periodic images. Spin-polarized calculations were performed and ferromagnetic states were assumed in the initial condition except for bcc Cr. The total energy in the self-consistent field (scf) calculations was converged within $10^{-6}$ Ry, and the total energy and forces in the geometry relaxation were converged within $10^{-4}$ Ry and $10^{-3}$ a.u., respectively.

We used atomic simulation environment (ASE) \cite{ase} to prepare the bulk structures of single-component systems and used pymatgen \cite{pymatgen} to extract binary and ternary systems in the cubic and hexagonal structures from the MP database \cite{materalsproject}. We excluded the crystals containing noble gas and actinides atoms. The ASE \cite{ase} was used to prepare slab models and the spglib \cite{spglib} was used to find a primitive cell for slabs. The VESTA was used to visualize crystal structures \cite{vesta}.  

\begin{figure}
\center\includegraphics[scale=0.5]{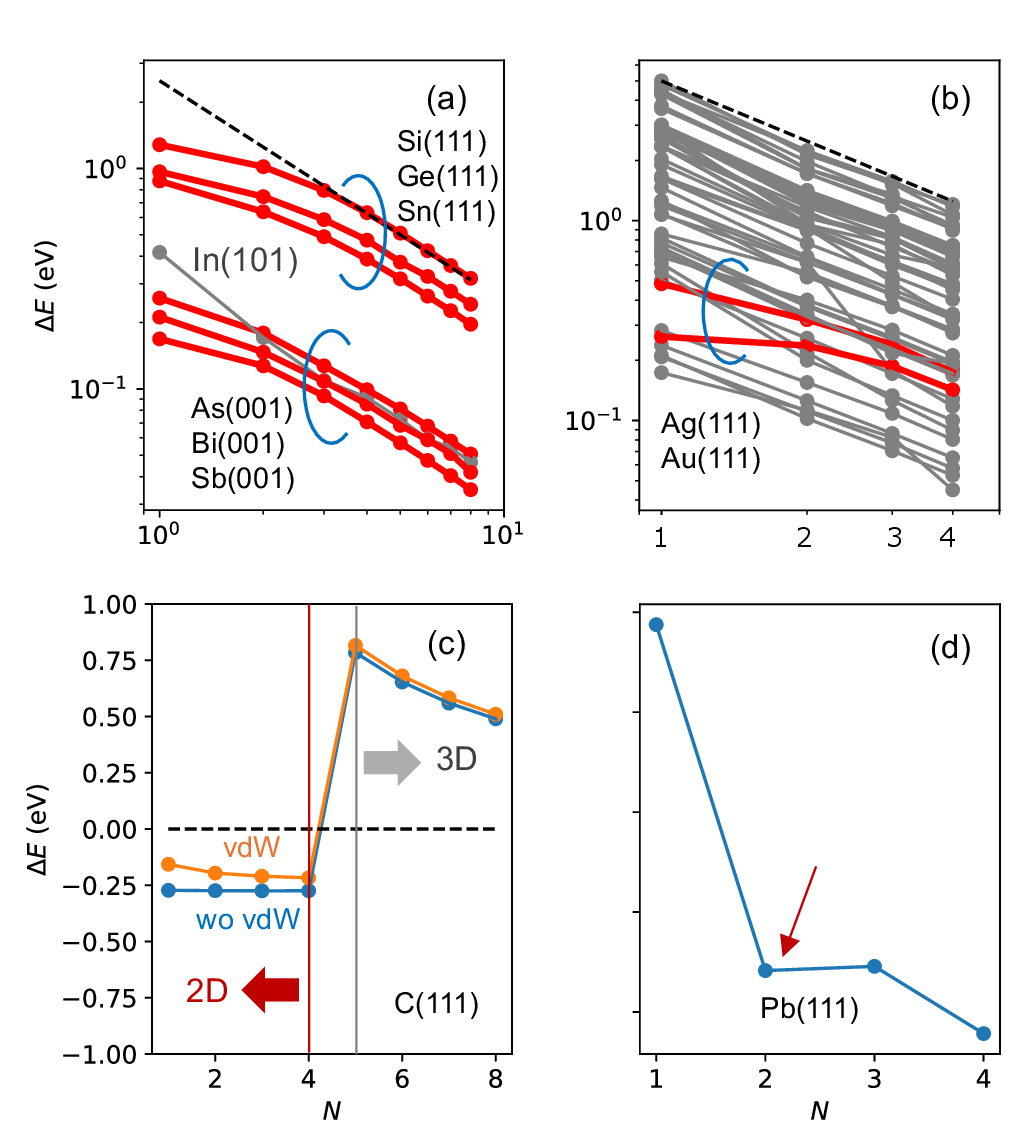}
\caption{$\Delta E$ versus $N$ for the single-component systems: (a) Si, Ge, Sn, As, Sb, Bi, and In, (b) fcc, bcc, and hcp metals, (c) C, and (d) Pb. The dashed line in log-log plots (a) and (b) indicates $\Delta E \propto N^{-1}$, and the curves with $p(1)>-0.6$ are colored in red. (c) The carbon diamond with and without the vdW correction exhibits $\Delta E<0$ for $N\le 4$, indicating a 3D-2D (diamond-graphene) transition at $N=4$. (d) $\Delta E$ of the Pb thin film has a minimal value at $N=2$, which corresponds to the plumbene.  } \label{fig1} 
\end{figure}

\begin{figure*}
\center\includegraphics[scale=0.53]{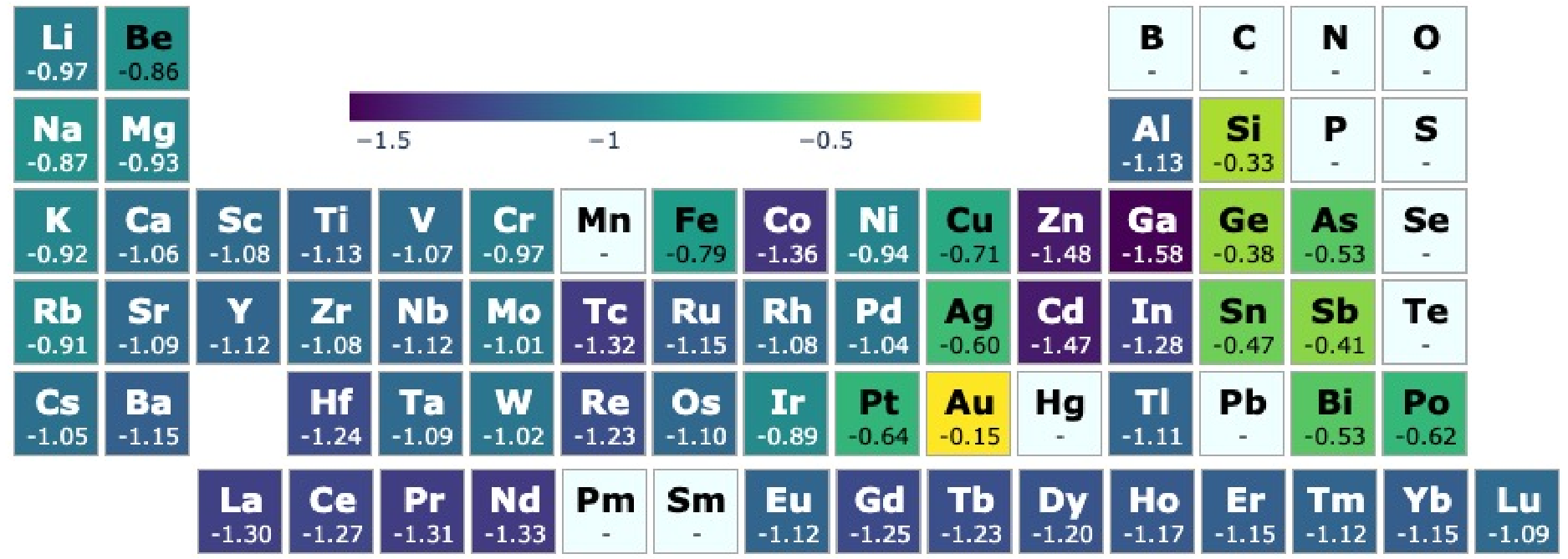}
\caption{$p(1)$ of the single-component systems. If $p(1)=-1$, the $N^{-1}$ law holds. The elements with negatively small $p(1)$ can be non-vdW 2D material. Mn (cubic), Pm, Sm (rhombohedral), and Hg (rhombohedral) are not studied due to their complex or unknown geometry. $p(1)$ of Pb is not shown because the $\Delta E$-$N$ curve is nonmonotonic (see Fig.~\ref{fig1}(d)). The corrections of vdW forces and SOC are included in As and Po, respectively. } \label{fig2} 
\end{figure*}

{\it Single-component system---}Figure \ref{fig1}(a) shows $\Delta E$ of $N$-layer thin films truncated from the (111) surface of Si, Ge, and Sn in the diamond structure, the (001) surface of As, Sb, and Bi in the rhombohedral structure, and the (101) surface of In in the orthorhombic structure. The thin films in the monolayer limit have the buckled honeycomb structure except for the In monolayer. The $\Delta E$ is inversely proportional to $N$ for large $N$, validating our theorem. The curves deviate from the $N^{-1}$ law at $N=1$ and 2 and $p(1)$ is larger than $-0.6$. This indicates that the thin films in the monolayer limit are no longer bulk and tend to change into 2D materials except for In. Our calculations suggest that a free-standing In monolayer does not exist, whereas the In monolayer may be supported by the SiC(0001) substrate \cite{2DIn}. Note also that the vdW correction to the exchange-correlation energy \cite{vdw3} was included in As because relatively small value of $\Delta E\simeq 10^{-3}$ eV independent of $N$ was observed within PBE. 

Figure \ref{fig1}(b) shows the $N$ dependence of $\Delta E$ for fcc, bcc, and hcp metals from Li to Pb. When truncating thin films, we considered nearly close packed surfaces, i.e., the (100) and (111) surfaces for fcc metals, the (100) and (110) surfaces for bcc metals, and the (0001) surface for hcp metals, because their surface energies tend to be smaller than the other surfaces. Then, the energetically stable thin films were adopted for fcc and bcc metals. Many metals follow the $N^{-1}$ law. Interestingly, Au(111) at $N=1$ (i.e., goldene) exhibits $p(1)=-0.15$, which is very close to zero. We also identified silverene (2D Ag). However, $p(1)$ is negatively large ($-0.6$), and the silverene has not been synthesized yet. 

The carbon should be studied separately because the diamond structure is metastable. As $N$ is decreased, $\Delta E$ becomes suddenly negative at $N=4$ (see Fig.~\ref{fig1}(c)). This indicates a 3D-2D transition. The thin films with $N\ge 5$ still form the diamond structure with the $sp^3$ bonds, while those with $N\le 4$ are already regarded as $N$-layered graphene with the $sp^2$ bonds. An inclusion of the vdW correction \cite{vdw3} also yields the 3D-2D transition at $N=4$ and a small increase in $\Delta E$ in the monolayer limit. 

It is interesting that a minimal $\Delta E$ is observed for the Pb thin film at $N=2$ [see Fig.~\ref{fig1}(d)]. Such a thin film relaxes to the buckled honeycomb structure, which is consistent with the crystal structure of plumbene \cite{yuhara}. 

In this way, the energy variation calculations of thin films enable us to identify the single-component non-vdW 2D materials that have been synthesized in experiments, validating our approach. The results are summarized in Fig.~\ref{fig2}. We apply our method to more complex systems below. 

Note that Po is known to crystalize into the simple cubic structure in bulk. The poloniumene in the square lattice structure (2D Po) truncated from the (001) surface has been predicted to be dynamically stable \cite{ono84}. The present calculation predicts $p(1)\simeq -0.6$ for 2D Po, which is close to that of Ag and Pt (see Fig.~\ref{fig2}), where the effect of spin-orbit coupling (SOC) is included in Po.  

\begin{figure}
\center\includegraphics[scale=0.4]{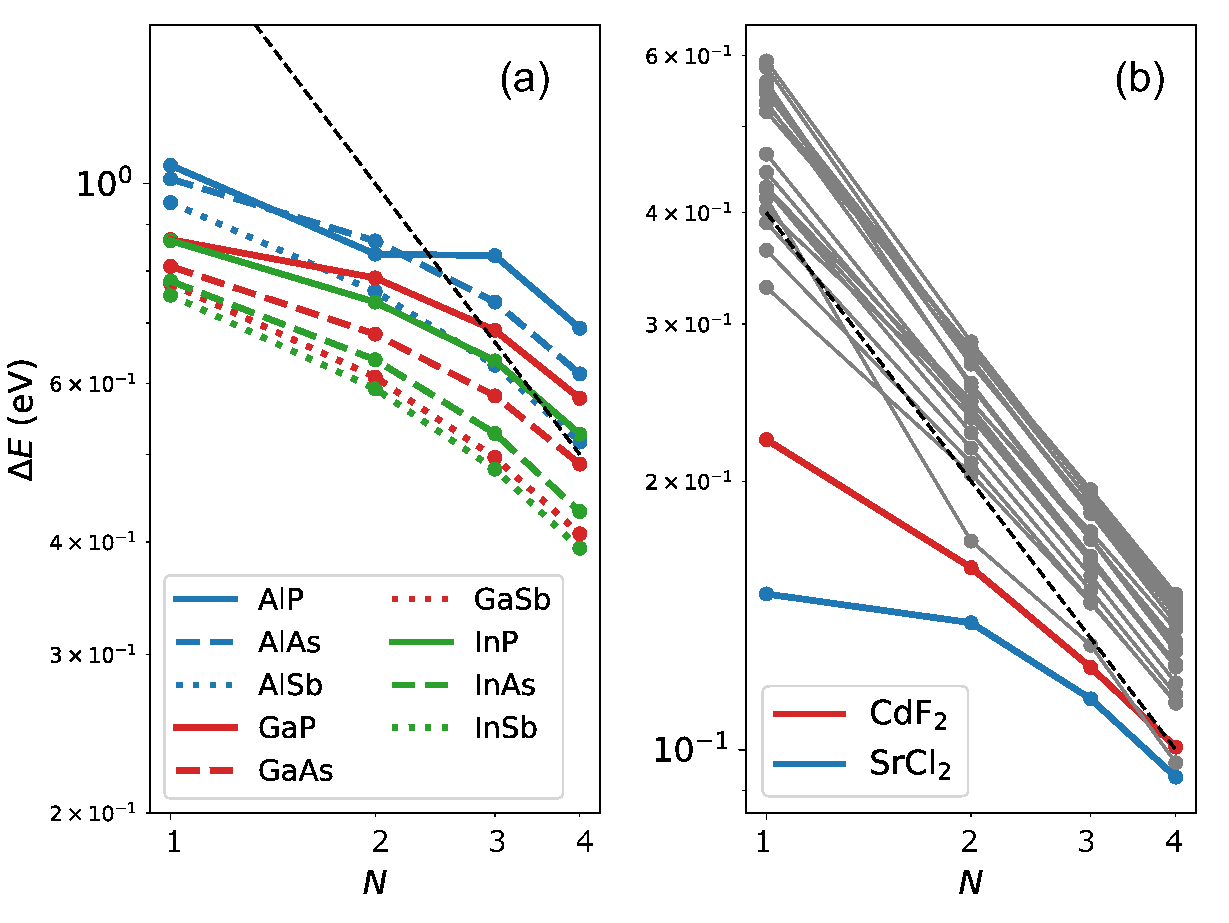}
\caption{The $N$-dependence of $\Delta E$ of thin films with the (111) surface for (a) III-V materials and (b) the fluorite-type materials. The dashed line indicates $\Delta E \propto N^{-1}$. All III-V materials exhibit $p(1)>-0.4$ and AlP has $p(2)\simeq 0$. If $p(1)>-0.5$, the $\Delta E$-$N$ curve is colored. } \label{fig3} 
\end{figure}

\begin{figure*}
\center\includegraphics[scale=0.6]{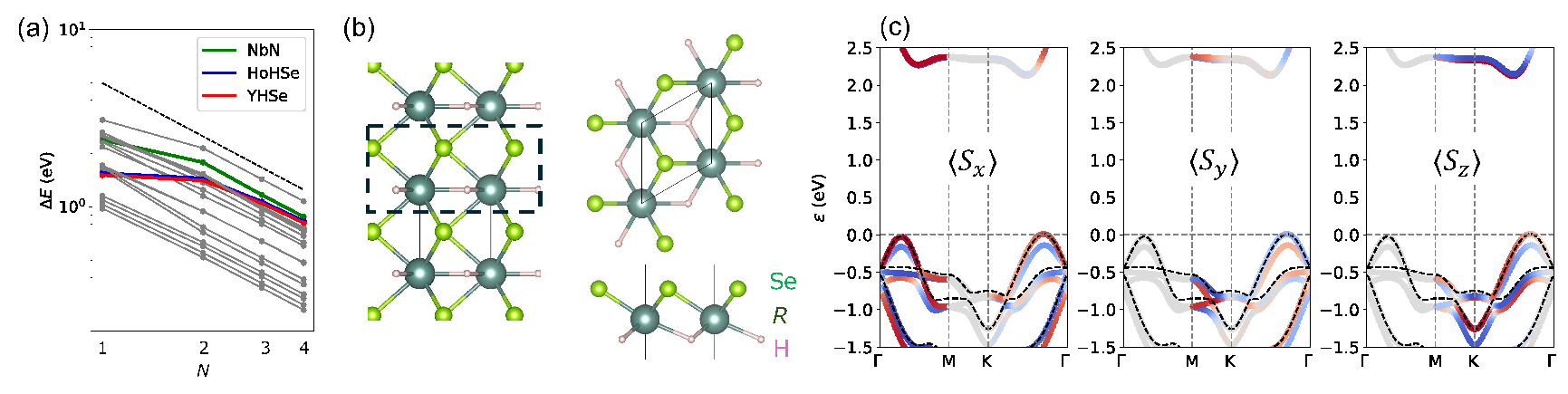}
\caption{(a) $\Delta E$ as a function of $N$ for hexagonal systems with $P\bar{6}m2$. YHSe (red) and HoHSe (blue). (b) Crystal structure of 3D $R$HSe (left) with $R=$Y and Ho. When the $N=1$ thin film is truncated (dashed), the monolayer relaxes to the Janus 1T structure (right). (c) The electronic band structure of YHSe monolayer with SOC. The expectation values of the spin operator on the spinor wave-functions are also plotted. The valence bands without SOC are also plotted (dashed). } \label{fig4} 
\end{figure*}

\begin{figure}
\center\includegraphics[scale=0.4]{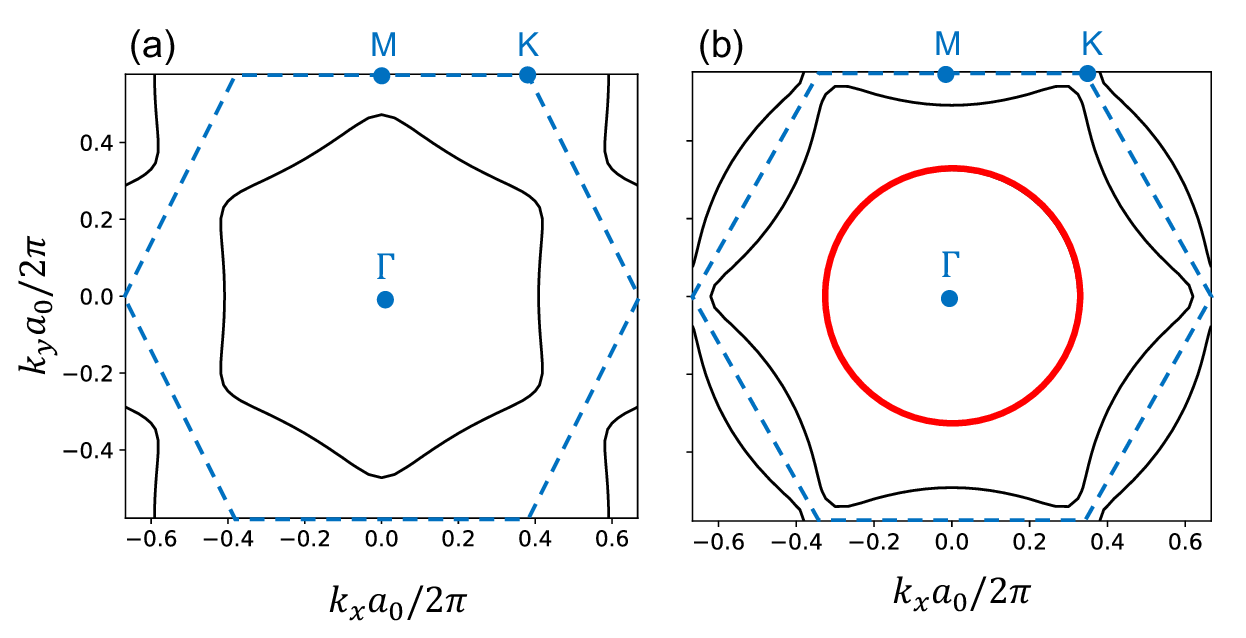}
\caption{The Fermi surface of (a) Au and (b) Al monolayers in the hexagonal structure. Dashed line indicates the boundary of the first Brillouin zone. The circular (red) and deformed hexagonal Fermi surfaces in Al monolayer exhibit out-of-plane and in-plane bond characters, respectively (see main text). } \label{fig5} 
\end{figure}

{\it Cubic crystal system---}Mounet {\it et al}. predicted that no vdW 2D materials are present in the cubic crystal system using high-throughput DFT calculations \cite{mounet}. However, several zincblende-type \cite{2Dtraditional} and fluorite-type \cite{ono2024} materials have been predicted to be non-vdW 2D materials. To reinvestigate what cubic crystals become non-vdW 2D materials, we apply our approach to these systems. In the present work, we calculate the total energy of thin films truncated from the (111) surface up to $N=4$. We considered III-V materials $AB$ ($A=$ Al, Ga, and In; $B=$ P, As, and Sb) and $AB_2$ and $A_2B$ fluorite-type semiconductors ($A$ indicates metallic elements and $B$ indicates chalcogen and halogen atoms) \cite{ono2024}. Their thin films at $N=1$ correspond to the buckled honeycomb and 1T structures, respectively. 

Figure \ref{fig3}(a) and \ref{fig3}(b) show $\Delta E$ as a function of $N$ for thin films of III-V materials and fluorite-type materials, respectively. All III-V materials have $p(1)>-0.4$ and exhibit a deviation from the $N^{-1}$ law. AlP shows $p(2)\simeq 0$. The crystal structure at $N=2$ is different from the double-layer honeycomb structure reported in Ref.~\cite{2Dtraditional}. Atomic displacements within a layer need to be assumed in the initial geometry to reproduce the latter. Still, the present calculations suggest that III-V materials can adopt several 2D structures. 

In Fig.~\ref{fig3}(b), a large deviation from the $N^{-1}$ law is observed in SrCl$_2$ and CdF$_2$ that have $p(1)=-0.11$ and $-0.48$, respectively. The previous study has reported that these 1T structures are dynamically stable and have a band gap of 5.76 eV and 3.85 eV (PBE), respectively \cite{ono2024}. We expect that the SrCl$_2$ monolayer is synthesized in future experiments, as $p(1)$ of SrCl$_2$ is close to that of goldene. 




{\it Hexagonal crystal system---}Many vdW 2D materials belonging to the hexagonal crystal system have been studied for twenty years. To explore non-vdW 2D materials with the hexagonal symmetry, we consider binary and ternary compounds with the space group of $P\bar{6}m2$. By using the MP database \cite{materalsproject}, we found 28 thermodynamically stable compounds including tungsten carbide-type (WC, $B_h$) and BaPtSb-type structures \cite{aflow2017}. 

We truncate thin films from the (0001) surface and calculate $\Delta E$ from $n=1$ to 4. We identified that $R$HSe ($R=$Y and Ho) shows a deviation from the $N^{-1}$ law (see Fig.~\ref{fig4}(a)). In the parent crystal, 2H-type $R$Se$_2$ layers are stacked along the $c$ axis but Se layers are shared with adjacent $R$Se$_2$ layers, and H atoms form the triangular lattice in the same plane of $R$ (see Fig.~\ref{fig4}(b), left). In the monolayer, we observe that H atoms move to bottom side of $R$ layer, and obtain Janus 1T-type monolayers, where the triangular lattice of $R$ is sandwiched by H and Se layers (see Fig.~\ref{fig4}(b), right). This is stark contrast with the existing Janus monolayers that are artificially created from transition metal dichalchogenides \cite{janus2017}. No imaginary frequencies were observed in their phonon calculations, showing that they are dynamically stable (see Supplemental Materials \cite{SM}). 

The Janus monolayers have no inversion symmetry, causing an internal electric field. Therefore, the spin degeneracy is lifted by including SOC \cite{yao,hu,chen2021}. Figure \ref{fig4}(c) shows the electronic band structure of 2D YHSe within PBE and PBE+SOC, showing an indirect band gap about 2 eV. With SOC, the spin-splitting is significant in the valence bands; the $S_x$ component dominates the spin polarization along the $\Gamma$-M line, while the $S_y$ and $S_z$ components are large along the $\Gamma$-K-M line. The spin-split bands due to SOC is also found in 2D HoHSe, which is provided in Supplemental Materials \cite{SM}. 

It is noteworthy that NbN monolayer was identified as a non-vdW 2D material (see Fig.~\ref{fig4}(a)). The physical properties and potential applications for spintronics have been studied in detail \cite{NbN2019,NbN2022}. 


{\it Discussion---}Before closing, we discuss a possible reason why goldene \cite{kashiwaya} is so unique and serves as a non-vdW 2D material. For single-component semiconductors classified as covalent crystals, the interatomic bonds can be changed into in-plane bonds, which is preferable for the formation of 2D materials. Therefore, it is not trivial how to stabilize goldene with metallic bonds. The goldene has a hexagon-shaped Fermi surface (see Fig.~\ref{fig5}(a)) and the electron states on each edge exhibit mainly in-plane bond character (20\% $6s$, 28\% $6p_x+6p_y$, and 52\% $5d_{xy}+5d_{x^2-y^2}$) \cite{ono2025}. The in-plane character allows the goldene to be robust against perturbations. On the other hand, aluminene (2D Al) has a circular Fermi surface with $p_z$ character and a deformed hexagon-shaped Fermi surface with $p_x$ and $p_y$ characters (see Fig.~\ref{fig5}(b)). The coexistence of $p_x$, $p_y$, and $p_z$ states on the Fermi surface is not preferable for stabilizing 2D flat sheet, giving rise to a buckling \cite{ono2020} or 3D crystallization by perturbations. 

{\it Conclusion---}In conclusion, we proposed that a downward deviation from the $N^{-1}$ law of exfoliation energy is an indicator for identifying non-vdW 2D materials. Several 2D materials that are absent from the existing databases have been revisited. We found that the Janus structures hidden in 3D crystals can appear in the monolayer limit with keeping the stoichometry. YHSe and HoHSe are representative materials that exhibit such a property. The values of $p(1)$ for the cubic and hexagonal crystals studied in this work are provided in Supplemental Materials \cite{SM}. 

This work can be extended by considering different parent crystal structures, surface cuts, and spin distribution, and using more advanced exchange-correlation functionals, in combination with high-throughput calculations. Predicting physical properties will further stimulate experimental syntheses of non-vdW 2D materials.  

\begin{acknowledgments}
This work was supported by JSPS KAKENHI (Grant No. 21K04628 and No. 24K01142). Calculations were done using the facilities of the Supercomputer Center, the Institute for Solid State Physics, the University of Tokyo.
\end{acknowledgments}




\begin{thebibliography}{99}

\bibitem{graphene} K. S. Novoselov, A. K. Geim, S. Morozov, D. Jiang, Y. Zhang, S. Dubonos, I. Grigorieva, and A. Firsov, Electric field effect in atomically thin carbon films, Science {\bf 306}, 666 (2004).

\bibitem{lebegue} S. Leb\`egue, T. Bj\"orkman, M. Klintenberg, R. M. Nieminen, and O. Eriksson, Two-Dimensional Materials from Data Filtering and Ab Initio Calculations, Phys. Rev. X {\bf 3}, 031002 (2013).

\bibitem{ashton} M. Ashton, J. Paul, S. B. Sinnott, and R. G. Hennig, Topology-Scaling Identification of Layered Solids and Stable Exfoliated 2D Materials, Phys. Rev. Lett. {\bf 118}, 106101 (2017).

\bibitem{mounet} N. Mounet, M. Gibertini, P. Schwaller, D. Campi, A. Merkys, A. Marrazzo, T. Sohier, I. E. Castelli, A. Cepellotti, G. Pizzi, and N. Marzari, Two-dimensional materials from high-throughput computational exfoliation of experimentally known compounds, Nat. Nanotech. {\bf 13}, 246 (2018). 

\bibitem{C2DB2021} M. N. Gjerding, A. Taghizadeh, A. Rasmussen, S. Ali, F. Bertoldo, T. Deilmann, U. P. Holguin, N. R. Kn{\o}sgaard, M. Kruse, A. H. Larsen {\it et al.}, Recent Progress of the Computational 2D Materials Database (C2DB), 2D Materials {\bf 8}, 044002 (2021). 

\bibitem{lyngby} P. Lyngby and K. S. Thygesen, Data-driven discovery of 2D materials by deep generative models, npj Comput Mater {\bf 8}, 232 (2022). 

\bibitem{hong} Y. L. Hong, Z. Liu, L. Wang, T. Zhou, W. Ma, C. Xu, S. Feng, L. Chen, M. L. Chen, D. M. Sun {\it et al.}, Chemical vapor deposition of layered two-dimensional MoSi$_2$N$_4$ materials, Science {\bf 369}, 670 (2020).




\bibitem{review} A. Molle, J. Yuhara, Y. Yamada-Takamura, and Z. Sofer, Synthesis of Xenes: physical and chemical methods, Chem. Soc. Rev. {\bf 54}, 1845 (2025). 

\bibitem{kashiwayareview} S. Kashiwaya, Y. Shi, J. Rosen, and L. Hultman, Perspectives on noble metallenes: from synthesis to application, 2D Mater. {\bf 12}, 033001 (2025). 

\bibitem{vogt} P. Vogt, P. De Padova, C. Quaresima, J. Avila, E. Frantzeskakis, M. C. Asensio, A. Resta, B. Ealet, and G. Le Lay, Silicene: Compelling Experimental Evidence for Graphenelike Two-Dimensional Silicon, Phys. Rev. Lett. {\bf 108}, 155501 (2012). 

\bibitem{yuhara} J. Yuhara, B. He, N. Matsunami, M. Nakatake, and G. Le Lay, Graphene's Latest Cousin: Plumbene Epitaxial Growth on a "Nano WaterCube", Adv. Mater. {\bf 31}, 1901017 (2019).

\bibitem{kashiwaya} S. Kashiwaya, Y. Shi, J. Lu, D. G. Sangiovanni, G. Greczynski, M. Magnuson, M. Andersson, J. Rosen, and L. Hultman, Synthesis of goldene comprising single-atom layer gold, Nat. Synth. {\bf 3}, 744 (2024).

\bibitem{ji2019} D. Ji, S. Cai, T. R. Paudel, H. Sun, C. Zhang, L. Han, Y. Wei, Y. Zang, M. Gu, Y. Zhang, W. Gao, H. Huyan, W. Guo, D. Wu, Z. Gu, E. Y. Tsymbal, P. Wang, Y. Nie, and X. Pan, Freestanding crystalline oxide perovskites down to the monolayer limit, Nature {\bf 570}, 87 (2019). 

\bibitem{puthirath2021} A. B. Puthirath, A. P. Balan, E. F. Oliveira, V. Sreepal, F. C. Robles Hernandez, G. Gao, N. Chakingal, L. M. Sassi, P. Thibeorchews, and G. Costin {\it et al.}, Apparent ferromagnetism in exfoliated ultrathin pyrite sheets, J. Phys. Chem. C {\bf 125}, 18927 (2021).

\bibitem{balan2022} A. P. Balan, A. B. Puthirath, S. Roy, G. Costin, E. F. Oliveira, M. A. S. R. Saadi, V. Sreepal, R. Friedrich, P. Serles, and A. Biswas {\it et al.}, Non-van der Waals quasi-2D materials; Recent advances in synthesis, emergent properties and applications, Mater. Today {\bf 58}, 164 (2022).


\bibitem{nevalaita2018} J. Nevalaita and P. Koskinen, Atlas for the properties of elemental two-dimensional metals, Phys. Rev. B {\bf 97}, 035411 (2018).

\bibitem{2Dtraditional} M. C. Lucking, W. Xie, D.-H. Choe, D. West, T.-M. Lu, and S. B. Zhang, Traditional Semiconductors in the Two-Dimensional Limit, Phys. Rev. Lett. {\bf 120}, 086101 (2018). 

\bibitem{ono2020} S. Ono, Dynamical stability of two-dimensional metals in the periodic table, Phys. Rev. B {\bf 102}, 165424 (2020).

\bibitem{pawar2022} A. A. Sangolkar, R. Agrawal, and R. Pawar, A prospectus for thickness dependent electronic properties of two-dimensional metals using density functional theory calculation, Int. J. Quantum Chem. {\bf 122}, e26982 (2022).

\bibitem{friedrich2022} R. Friedrich, M. Ghorbani-Asl, S. Curtarolo, and A. V. Krasheninnikov, Data-Driven Quest for Two-Dimensional Non-van der Waals Materials, Nano Lett. {\bf 22}, 989 (2022). 

\bibitem{ono2024} S. Ono and R. Pawar, Fluorite-type materials in the monolayer limit, Phys. Rev. Materials {\bf 8}, 094002 (2024).

\bibitem{pereira2025} M. L. Pereira Jr, E. J. A. dos Santos, L. A. Ribeiro Jr, and D. S. Galvao, How does goldene stack?, Mater. Horiz. {\bf 12}, 1144 (2025). 



\bibitem{ono2021} S. Ono and H. Satomi, High-throughput computational search for two-dimensional binary compounds: Energetic stability versus synthesizability of three-dimensional counterparts, Phys. Rev. B {\bf 103}, L121403 (2021).

\bibitem{sun} W. Sun, S. T. Dacek, S. P. Ong, G. Hautier, A. Jain, W. D. Richards, A. C. Gamst, K. A. Persson, and G. Ceder, The thermodynamic scale of inorganic crystalline metastability, Sci. Adv. {\bf 2}, e1600225 (2016).

\bibitem{de} D. S. De, B. Schaefer, B. von Issendorff, and S. Goedecker, Nonexistence of the decahedral Si$_20$H$_20$ cage: Levinthal's paradox revisited, Phys. Rev. B {\bf 101}, 214303 (2020).


\bibitem{pakdel} S. Pakdel, T. Olsen, and K. S. Thygesen, Effect of Hubbard U-corrections on the electronic and magnetic properties of 2D materials: a high-throughput, npj Comput Mater {\bf 11}, 18 (2025).

\bibitem{wines} D. Wines, K. Choudhary, A. J. Biacchi, K. F. Garrity, and F. Tavazza, High-Throughput DFT-Based Discovery of Next Generation Two-Dimensional (2D) Superconductors, Nano Lett. {\bf 23}, 969 (2023). 

\bibitem{jang} B. G. Jang, C. Lee, J.-X. Zhu, and J. H. Shim, Exploring two-dimensional van der Waals heavy-fermion material: Data mining theoretical approach, npj 2D Mater Appl {\bf 6}, 80 (2022).

\bibitem{wang} V. Wang, G. Tang, Y.-C. Liu, R.-T. Wang, H. Mizuseki, Y. Kawazoe, J. Nara, and W. T. Geng, High-Throughput Computational Screening of Two-Dimensional Semiconductors, J. Phys. Chem. Lett. {\bf 13}, 11581 (2022). 

\bibitem{zhang} C. Zhang, R. Wang, H. Mishra, and Y. Liu, Two-Dimensional Semiconductors with High Intrinsic Carrier Mobility at Room Temperature, Phys. Rev. Lett. {\bf 130}, 087001 (2023). 


\bibitem{qe} P. Giannozzi, O. Andreussi, T. Brumme, O. Bunau, M. B. Nardelli, M. Calandra, R. Car, C. Cavazzoni, D. Ceresoli, and M. Cococcioni {\it et al}., Advanced capabilities for materials modeling with Quantum ESPRESSO, J. Phys.: Condens. Matter {\bf 29}, 465901 (2017).

\bibitem{pbe} J. P. Perdew, K. Burke, and M. Ernzerhof, Generalized Gradient Approximation Made Simple, Phys. Rev. Lett. {\bf 77}, 3865 (1996).

\bibitem{dalcorso} A. Dal Corso, Pseudopotentials periodic table: From H to Pu, Computational Material Science {\bf 95}, 337 (2014).


\bibitem{smearingMV} N. Marzari, D. Vanderbilt, A. De Vita, and M. C. Payne, Thermal contraction and disordering of the Al(110) surface, Phys. Rev. Lett. {\bf 82}, 3296 (1999).



\bibitem{ase} A. H. Larsen, J. J. Mortensen, J. Blomqvist, I. E. Castelli, R. Christensen, M. Du{\l}ak, J. Friis, M. N. Groves, B. Hammer, C. Hargus {\it et al.}, The atomic simulation environment---a Python library for working with atoms, J. Phys.: Condens. Matter {\bf 29}, 273002 (2017). 

\bibitem{pymatgen} S. P. Ong, W. D. Richards, A. Jain, G. Hautier, M. Kocher, S. Cholia, D. Gunter, V. L. Chevrier, K. A. Persson, and G. Ceder, Python Materials Genomics (pymatgen): A robust, open-source python library for materials analysis, Comput. Mater. Sci. {\bf 68}, 314 (2013).

\bibitem{materalsproject} A. Jain, S. P. Ong, G. Hautier, W. Chen, W. D. Richards, S. Dacek, S. Cholia, D. Gunter, D. Skinner, G. Ceder, and K. A. Persson, The Materials Project: A materials genome approach to accelerating materials innovation, APL Mater. {\bf 1}, 011002 (2013).

\bibitem{spglib} A. Togo, K. Shinohara, and I. Tanaka, Spglib: a software library for crystal symmetry search, Sci. Technol. Adv. Mate. {\bf 4}, 2384822 (2024).

\bibitem{vesta} K. Momma and F. Izumi, VESTA 3 for three-dimensional visualization of crystal, volumetric and morphology data, J. Appl. Crystallogr. {\bf 44}, 1272 (2011). 


\bibitem{2DIn} M. Bauernfeind, J. Erhardt, P. Eck, P. K. Thakur, J. Gabel, T.-L. Lee, J. Sch\"{a}fer, S. Moser, D. Di Sante, R. Claessen, and G. Sangiovanni, Design and realization of topological Dirac fermions on a triangular lattice, Nat. Commun. {\bf 12}, 5396 (2021).

\bibitem{vdw3} S. Grimme, J. Antony, S. Ehrlich, and H. Krieg, A consistent and accurate ab initio parametrization of density functional dispersion correction (DFT-D) for the 94 elements H-Pu, J. Chem. Phys {\bf 132}, 154104 (2010). 


\bibitem{ono84} S. Ono, Two-dimensional square lattice polonium stabilized by the spin-orbit coupling, Sci. Rep. {\bf 10}, 11810 (2020). 










\bibitem{aflow2017} M. J. Mehl, D. Hicks, C. Toher, O. Levy, R. M. Hanson, G. L. W. Hart, and S. Curtarolo, The AFLOW Library of Crystallographic Prototypes: Part 1, Comp. Mat. Sci. {\bf 136}, S1 (2017).


\bibitem{janus2017} A. Y. Lu, H. Zhu, J. Xiao, C. P. Chuu, Y. Han, M. H. Chiu, C. C. Cheng, C. W. Yang, K. H. Wei, Y. Yang {\it et al.}, Janus monolayers of transition metal dichalcogenides. Nature Nanotech. {\bf 12}, 744 (2017).


\bibitem{SM} See Supplemental Material at (URL) for (i) the values of $p(1)$ of zincblende, fluorite, WC, and BaPtSb-type structures, (ii) phonon dispersions of YHSe and HoHSe monolayers calculated by using density-functional perturbation theory \cite{dfpt}, and (iii) electronic band structure of HoHSe within PBE+SOC.   

\bibitem{dfpt} S. Baroni, S. Gironcoli, A. Dal Corso, and P. Giannozzi, Phonons and related crystal properties from density-functional perturbation theory, Rev. Mod. Phys. {\bf 73}, 515 (2001).



\bibitem{yao} Q. F. Yao, J. Cai, W. Y. Tong, S. J. Gong, J. Q. Wang, X. Wan, C. G. Duan, and J. H. Chu, Manipulation of the large Rashba spin splitting in polar two-dimensional transition-metal dichalcogenides, Phys. Rev. B {\bf 95}, 165401 (2017). 

\bibitem{hu} T. Hu, F. Jia, G. Zhao, J. Wu, A. Stroppa, and W. Ren, Intrinsic and anisotropic Rashba spin splitting in Janus transition-metal dichalcogenide monolayers, Phys. Rev. B {\bf 97}, 235404 (2018). 


\bibitem{chen2021} J. Chen, K. Wu, W. Hu, and J. Yang, Spin-Orbit Coupling in 2D Semiconductors: A Theoretical Perspective, J. Phys. Chem. Lett. {\bf 12}, 12256 (2021). 

\bibitem{NbN2019} A. Chanana and U. V. Waghmare, Prediction of Coupled Electronic and Phononic Ferroelectricity in Strained 2D h-NbN: First-Principles Theoretical Analysis, Phys. Rev. Lett. {\bf 123}, 037601 (2019). 

\bibitem{NbN2022} R. Ahammed and A. De Sarkar, Valley spin polarization in two-dimensional $h$-$M$N ($M=$Nb,Ta) monolayers: Merger of valleytronics with spintronics, Phys. Rev. B {\bf 105}, 045426 (2022). 


\bibitem{ono2025} S. Ono and H. Yoshioka, Anomalous chirality dependence of strain energy in gold nanotubes, Phys. Rev. B {\bf 111}, 085414 (2025). 


\end{thebibliography}

\end{document}